\documentclass[a4paper,superscriptaddress]{jpconf}

\usepackage{graphicx}

\begin{document}
\title{Conjecture on reflectionlessness of blood-vascular system as a wave-conducting medium}

\author{Denis S Goldobin$^{1,2}$ and Irina A Mizeva$^{1}$}
\address{$^1$Institute of Continuous Media Mechanics UB RAS,
         1 Akademika Koroleva street, 614013~Perm,~Russia}
\address{$^2$Department of Theoretical Physics, Perm State University,
         15 Bukireva street, 614990~Perm,~Russia}
\ead{Denis.Goldobin@gmail.com, mizeva@icmm.ru}

\begin{abstract}
Our research is related to the employment of photoplethysmography (PPG) and laser Doppler flowmetry (LDF) techniques (measuring the blood volume and flux, respectively) for the peripheral vascular system. We derive the governing equations of the wave dynamics for the case of extremely inhomogeneous parameters. We argue for the conjecture that the blood-vascular system as a wave-conducting medium should be nearly reflection-free. With the reflectionlessness condition, one can find the general solution to the governing equation and, on the basis of this solution, analyse the relationships between PPG- and LDF-signals.
\end{abstract}

\section{Introduction}
\begin{figure}
\begin{center}
\noindent
{\sf (a)}\hspace{-17pt}
\includegraphics[width=0.46\textwidth]%
  {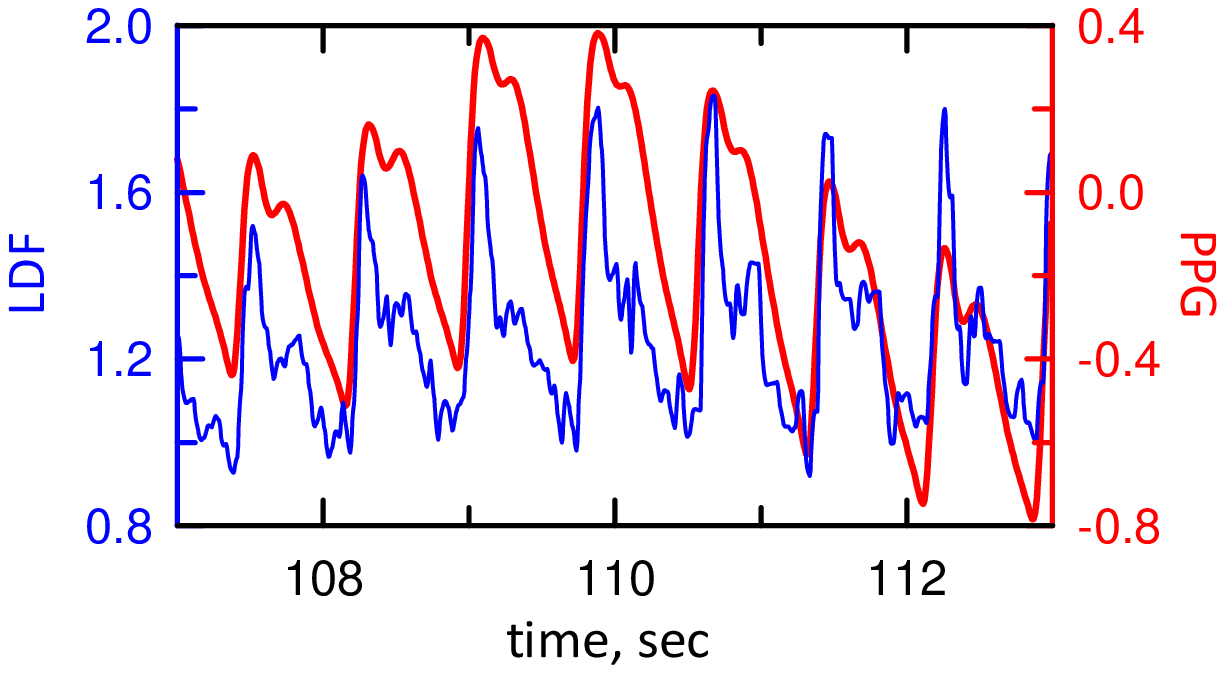}
\qquad
{\sf (b)}\hspace{-17pt}
\includegraphics[width=0.46\textwidth]%
  {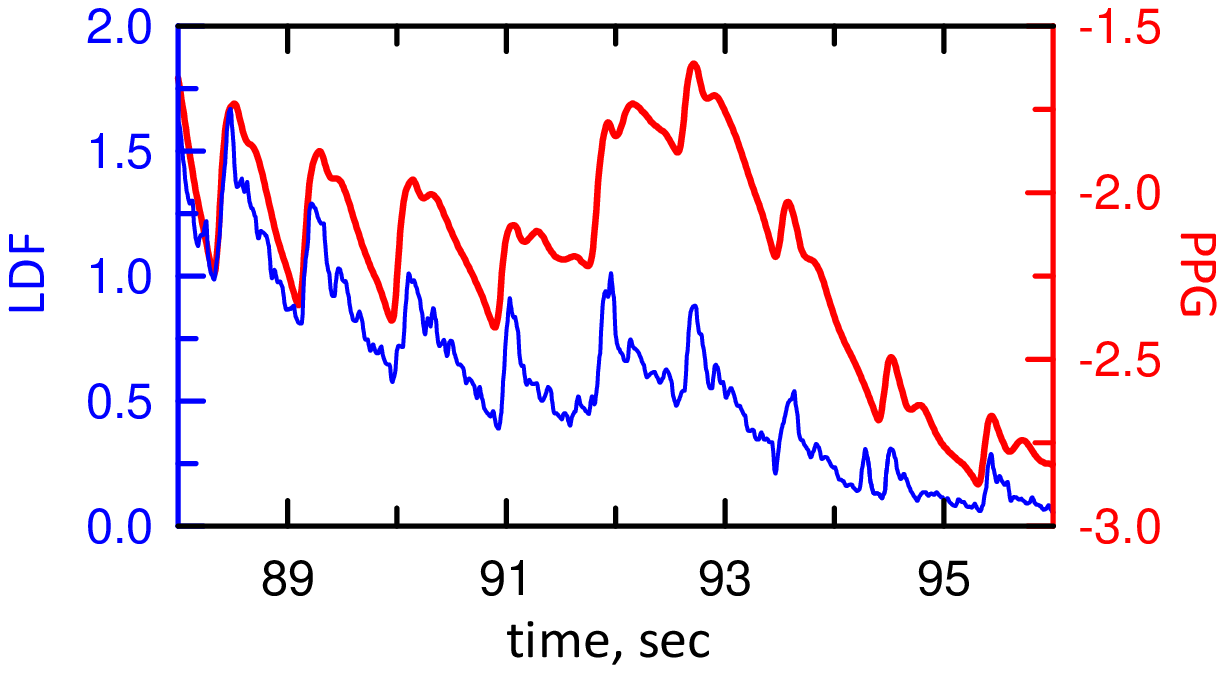}
\end{center}
\caption{Sample readings from the plethysmograph measuring the blood volume in vessels
(PPG, red thick line) and from the Doppler measuring the product of the flow velocity and the blood volume (LDF, blue thin line) are presented for the normal conditions (a) and for the functional testing conditions (b)}
\label{fig1}
\end{figure}

Currently, the filtration of blood through microvessels and tissues of the human body is a subject of growing interest of researchers. Two of the most popular noninvasive tools for experimental studies on the subject are photoplethysmography (PPG) and the laser Doppler flowmetry (LDF)~\cite{Allen-Frame-Murray-2002}. The PPG signal is proportional to the volume of blood in microvessels, while the LDF signal is proportional to the flow velocity multiplied by the number of erythrocytes in the scanned tissue volume. In figure~\ref{fig1}, one can see sample signals measured on human hand fingers. In particular, for a deep inspiratory gasp manoeuvre~\cite{Laude-etal-1993,Marco-etal-2012}, experimental observations reveal that, under certain conditions maintained during functional tests, a phase shift between the $1\,\mathrm{Hz}$-components of PPG- and LDF-signals appears~\cite{Podtaev-Mizeva-Alan-2012}, which illustrates how combined implementation of these two tools may be seminal for diagnostics of the state of the blood-vascular system. It is therefore imperative to improve our understanding of the relationships between PPG- and LDF-signals and of their dependance on the state of the vessels.

In this paper we focus on the consideration of the propagation of the waves of flux/pressure through the blood-vascular system as a wave-conducting medium with highly inhomogeneous properties. Specifically, the net cross-section area of vessels varies from $4.5$ to $200\,\mathrm{cm}^2$ along the system, which is as much as 2 orders of magnitude. With such a strong inhomogeneity of parameters, one can, as an approximation, neglect minor effects such as viscoelasticity of the vessel walls~\cite{Formaggia-Lamponi-Quarteroni-2003,Valdez-Jasso-etal-2009,Valdez-Jasso-etal-2011} (the viscoelasticity effect is treated as ``minor'' here only against the background of a very strong effect of inhomogeneity) and consider the walls to be merely elastic.

Further, for the time scale of a heart pulse $T=0.2\,\mathrm{s}$ and the characteristic value of blood kinematic viscosity $\nu=4\cdot10^{-6}\,\mathrm{m^2/s}$ [\footnote{As the blood is a non-Newtonian liquid, one can speak of a reference value of the viscosity coefficient for estimates but not of a rigorously defined coefficient.}], one can evaluate the thickness of the viscous boundary layer: $\delta_\mathrm{vis}=\sqrt{\nu T/\pi}\approx0.5\,\mathrm{mm}$. Hence, for the vessels of inner diameter $d_\mathrm{in}>1\,\mathrm{mm}$, one can assume an inviscid blood flow. In this paper we will restrict our consideration to the propagation of waves through the part of the blood-vascular system, where the vessel inner diameter $d_\mathrm{in}>1\,\mathrm{mm}$. For the thinner vessels the system is significantly more stiff (which is known from experiments and will be also apparent with the results of our derivation at the end of the paper) and the wave propagation through this remaining part is very fast. With this fast propagation of waves and the conditions of the mass conservation and the flux conservation, one can approximately assume the blood flux and volume to be nearly immediately transferred from the edge of the part of the blood-vascular system where $d_\mathrm{in}>1\,\mathrm{mm}$ to the microvascular system. To summarise, we will consider an inviscid blood flow and associate the blood flux and vessel volume oscillations at the distant-from-heart end with measured LDF- and PPG-signals.

Generally, in a wave-conducting medium, the inhomogeneity of parameters produces reflection of waves. The absence of reflection is possible but requires mutually consistent variation of inhomogenous parameters or their specific dependence on coordinates~\cite{Didenkulova-Pelinovsky-2009,Didenkulova-Pelinovsky-2012,Petrukhin-Pelinovsky-Batsyna-2011,Petrukhin-Pelinovsky-Batsyna-2012}. Moreover, a waveguide with as strong inhomogeneity of parameters as in the blood-vascular system would be generally nearly impenetrable for waves, reflecting the major part of the energy of incident waves. Meanwhile, in practice, one observes reflection of pulses from the aorta bifurcation point and similar reflections, but they are not as strong as could generally be. This suggests a plausible conjecture that in a healthy state the system is optimised in a way to minimise the reflections. The latter is actually wide-spread for natural systems with different sorts of adaptation dynamics~\cite{Didenkulova-Pelinovsky-2009,Didenkulova-Pelinovsky-2012}. The approximation of the reflectionlessness of the system will provide us with the opportunity to derive the general solution for the equations of wavy dynamics we will derive in this paper.

The paper is organised as follows. In Sec.\ 2, we consider the possibility of a phase shift between PPG- and LDF-signals for the stationary-profile waves without any restrictions on mechanical properties of the vessel walls.
In Sec.\ 3.1, we derive governing equations for the wave dynamics in the case of strong inhomogeneity of the system parameters. In Sec.\ 3.2, the general solution to these equations is obtained for the reflection-free case. In Secs.\ 3.4 and 3.5, on the basis of this solution, the relationships between the waves of the blood volume (PPG) and the flux (LDF) are established. Finally, in Sec.\ 4, we summarise our findings.

\section{Absence of the phase shift for stationary-profile waves}
Prior to considering the relationships between the waves of volume and flux for the general case of inhomogeneous parameters, it is suitable to answer a preliminary question: Why the phase shift is NOT to be considered as typical?

One can show that in an elastic tube with laminar flow of inviscid liquid the linear (small-amplitude) waves of pressure {\it or} widening of the tube propagate without dispersion and are governed by an equation of the type of the string vibration equation~\cite{Pedley-1980}.

\begin{figure}
\begin{center}
\includegraphics[width=0.45\textwidth]%
  {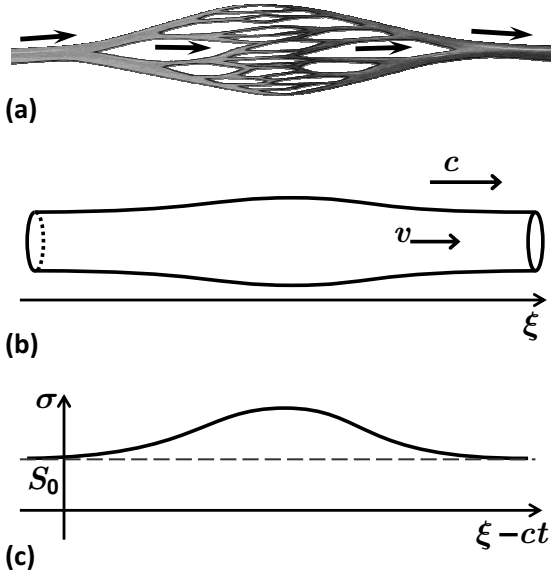}
\qquad
\begin{minipage}[b]{2.65in}
\caption{(a)~Sketch of the branching vascular network,
(b)~wave of deformation of an elastic tube containing liquid,
(c)~schematic picture of a volume wave propagating without dispersion}
\end{minipage}
\end{center}
\label{fig2}
\end{figure}

Let us consider the relationship between the wave of volume of an elastic tube and the wave of flux in it, without imposing any additional assumptions except for the stationarity of wave profile. We make a natural choice for the spatial coordinate; we use the coordinate $\xi$ which measures the length along the unstretched vessel. This coordinate is tied to the vessel wall, {\it i.e.} it becomes stretched in space when the vessel is stretched. To characterise both widening and stretching of a vessel we introduce the following parameters: $S$ which is the cross-section area of the vessel and
\[
\sigma=\frac{\delta V}{\delta\xi}
\]
which measures the volume per elementary length $\delta\xi$ of the vessel. Notice, $\sigma=S$ in the case of widening of a vessel without longitudinal stretching---for a stretched state $\sigma$ and $S$ are different. The average over cross-section velocity of the blood flow in the vessel will be denoted by $v$, and the wave propagation speed will be $c$ (see figure~\ref{fig2}b).
Accordingly, the liquid flux
\[
q=v\,S\,.
\]
Note, LDF measures the flow velocity multiplied by the number of erythrocytes, which is proportional to
\[
q_\mathrm{_{LDF}}=v\,\sigma
\]
and generally differs from the flux $q=v\,S$. However, specifically for the site of measurements in~\cite{Podtaev-Mizeva-Alan-2012}, one can neglect the elongation of capillaries and associate LDF-readings with flux $q$. In this paper we associate LDF for the microcirculatory system with the flux.

As the liquid is incompressible, one can write the condition of the volume conservation:
\begin{equation}
\frac{\partial}{\partial t}\sigma(\xi,t)=
-\frac{\partial}{\partial\xi}\Big(v(\xi,t)\,S(\xi,t)\Big)\,.
\label{eq21}
\end{equation}
We assume the wave propagation speed $c$ to be constant along the vessel and consider the volume wave of stationary profile (figure~\ref{fig2}c)
\begin{equation}
\sigma(\xi,t)=\sigma(\xi-ct)\,.
\label{eq22}
\end{equation}
For such a wave $(\partial/\partial t)=-c(\partial/\partial\xi)$.
Then equation~(\ref{eq21}) can be integrated with respect to $\xi$;
\begin{equation}
c\,(\sigma-\sigma_0)=q-q_0\,,
\label{eq23}
\end{equation}
where $\sigma_0$ and $q_0$ correspond to the unperturbed state of the vessel and the flux in it. As the vessel is not stretched in the unperturbed state, $\sigma_0=S_0$ and equation~(\ref{eq23}) can be recast as
\begin{equation}
q=c\sigma-cS_0+q_0\,.
\label{eq24}
\end{equation}
According to the latter equation, the flux and the vessel volume vary in-phase.

\section{Wave propagation through an inhomogeneous vascular network}
As shown in the previous section, in a vessel with homogeneous properties, the phase shift is impossible. Nonetheless, the question whether the phase shift can appear due to variation of parameters (elasticity and inner diameter) along the vessel or not remains unanswered. The construction of mathematical description of the system with inhomogeneous parameters is our task in the current section.

\subsection{Vessel with inhomogeneous parameters}
Equation~(\ref{eq21}) describes the liquid incompressibility and holds valid for the case where the diameter and the elasticity of the vessel vary along its length. The time-derivative of this equation yields
\begin{equation}
\frac{\partial^2}{\partial t^2}\sigma(\xi,t)=
-\frac{\partial}{\partial\xi}\left[\frac{\partial}{\partial t}\Big(v(\xi,t)\,S(\xi,t)\Big)\right].
\label{eq31}
\end{equation}
We consider perturbations against the background of a static state of a vessel with a time-independent flow;
 $\sigma(\xi,t)=\sigma_0(\xi)+\sigma_1(\xi,t)$, $S(\xi,t)=S_0(\xi)+S_1(\xi,t)$,
 $v(\xi,t)=v_0(\xi)+v_1(\xi,t)$ and pressure $p(\xi,t)=p_0(\xi)+p_1(\xi,t)$,
which can be used for calculation of the time-derivative in the r.h.s.\ of equation~(\ref{eq31}). The Bernoulli's equation for unsteady potential flow
\[
\frac{\partial v}{\partial t}=-\nabla\left(\frac{1}{\rho}p+\frac{v^2}{2}\right)
\]
yields for perturbations
\begin{equation}
\frac{\partial v_1}{\partial t}=-\nabla\left(\frac{1}{\rho}p_1+v_0v_1\right).
\label{eq32}
\end{equation}
Then the time-derivative
\begin{eqnarray}
\frac{\partial}{\partial t}(vS)=v_0(\xi)\frac{\partial}{\partial t}S_1(\xi,t)
 +S_0(\xi)\frac{\partial}{\partial t}v_1(\xi,t)
=v_0 \frac{\partial }{\partial t}S_1 -S_0 \nabla \left( {\frac{1}{\rho }p_1
+v_0 v_1 } \right)
\nonumber
\\[10pt]
{}
=v_0 \left( {\frac{\delta S}{\delta \sigma }} \right)\frac{\partial
}{\partial t}\sigma _1 -S_0 \nabla \left( {\frac{1}{\rho }\left(
{\frac{\delta \sigma }{\delta p}} \right)^{-1}\sigma _1 +v_0 v_1 } \right).
\qquad\qquad\qquad
\label{eq33}
\end{eqnarray}
The ratio of variations $(\delta S/\delta\sigma)$ under varying pressure in the vessel is a dimensionless quantity of the order of the magnitude of $1$ and depends on the ratio of the elasticity moduli and their possible anisotropy.

Let us estimate the orders of the magnitude of three contributions in
 $(\partial/\partial t)(vS)$
in equation~(\ref{eq33}). Fields $\sigma_k$, $S_k$, $v_k$, $p_k$ (where $k=0,1$) are of the same order of magnitude as in the case of a homogeneous vessel. From equation~(\ref{eq23}), one finds
 $c\sigma_1\sim v_0S_1+S_0v_1$.
As $S_1\sim\sigma_1$ and $v_0\ll c$, the velocity perturbation $v_1\approx c\sigma_1/S_0$.
For wave fields, $(\partial/\partial t)\sim-c(\partial/\partial\xi)$. Hence, three contributions in $(\partial/\partial t)(vS)$ are of the following orders of magnitude:
\[
v_0\frac{\partial\sigma_1}{\partial t}\,,
\qquad
\frac{S_0}{\rho c}\left(\frac{\delta\sigma}{\delta p}\right)^{-1}\frac{\partial\sigma_1}{\partial t}\,,
\qquad
v_0\frac{\partial\sigma_1}{\partial t}\,,
\]
in the order of their appearance. For a {\it vessel with homogeneous parameters} the relation
 $(\partial/\partial t)=-c(\partial/\partial\xi)$
is accurate, and from equations~(\ref{eq31}) and (\ref{eq33}) for $v_0\ll c$ one can obtain
\begin{equation}
c^2\approx\frac{S_0}{\rho}\left(\frac{\delta\sigma}{\delta p}\right)^{-1}.
\label{eq34}
\end{equation}
Thus, the second term in sum (\ref{eq33}) is of the order of magnitude of
 $c(\partial\sigma_1/\partial t)$
and large compared to the first and third terms, which $\sim v_0(\partial\sigma_1/\partial t)$.

Taking only the leading term of (\ref{eq33}) into account, one can recast equation~(\ref{eq31}) as
\begin{equation}
\frac{\partial^2}{\partial t^2}\sigma_1=\frac{\partial}{\partial\xi}
 \left[S_0(\xi)\frac{\partial}{\partial\xi}\left(\beta(\xi)\,\sigma_1\right)\right],
\label{eq35}
\end{equation}
where $\beta:=[\rho(\delta\sigma/\delta p)]^{-1}$ characterises the elasticity of the vessel walls.

Noteworthy, with proper functions $S_0(\xi)$ and $\beta(\xi)$, equation~(\ref{eq35}) is valid for an approximate description of propagation of waves through a bunch of parallel vessels with hierarchy of branchings.

\subsection{Solutions of equation~(\ref{eq35}) in the reflection-free case}
Equation~(\ref{eq35}) admits reflectionless wave propagation, if it can be recast in the form
\[
\frac{\partial}{\partial\zeta_1}\frac{\partial}{\partial\zeta_2}F=0\,.
\]
Introducing operator
\[
\hat{Q}F:=\frac{S_0(\xi)}{\alpha(\xi)}\frac{\partial}{\partial\xi}(\alpha(\xi)\,F)\,,
\]
one can rewrite equation~(\ref{eq35}) in the form
\[
\left(\frac{S_0(\xi)}{\beta(\xi)}\frac{\partial^2}{\partial t^2}-\hat{Q}^2\right)
 \frac{\beta(\xi)\sigma_1}{\alpha(\xi)}=0\,.
\]
The latter equation can be factorised,
\begin{equation}
\left(\sqrt{\frac{S_0(\xi)}{\beta(\xi)}}\frac{\partial}{\partial t}-\hat{Q}\right)
\left(\sqrt{\frac{S_0(\xi)}{\beta(\xi)}}\frac{\partial}{\partial t}+\hat{Q}\right)F=0\,,
\label{eq36}
\end{equation}
if
\begin{equation}
\frac{\partial}{\partial\xi}\sqrt{S_0(\xi)/\beta(\xi)}=0\,.
\label{eq37}
\end{equation}

Equation~(\ref{eq37}) is the condition of the reflectionlessness of the system. We keep the square root in this equation on purpose, since in reality the ratio $(S_0/\beta)$ is not perfectly constant along the vessel network and with equation~(\ref{eq37}) one can have a quantitative estimate how accurate the assumption of the reflectionlessness is.

For the waves of our interest, equation~(\ref{eq36}) yields
\[
\left(\frac{\partial}{\partial t}+\sqrt{\beta(\xi)\,S_0(\xi)}
\frac{\partial}{\partial\xi}\right)\beta(\xi)\,\sigma_1(\xi,t)=0\,,
\]
and thus
\begin{equation}
\sigma_1(\xi,t)=\frac{1}{\beta(\xi)}f\left(
 \int^\xi\frac{d\xi_1}{\sqrt{\beta(\xi_1)\,S_0(\xi_1)}} -t\right).
\label{eq38}
\end{equation}
The wave travel time for $\xi_0$ to $\xi$ is
\begin{equation}
\tau(\xi,\xi_0)=\int_{\xi_0}^\xi\frac{d\xi_1}{\sqrt{\beta(\xi_1)\,S_0(\xi_1)}}\,.
\label{eq39}
\end{equation}

Expression (\ref{eq38}) is a solution to equation~(\ref{eq35}) for arbitrary continuous function $f$. From equation~(\ref{eq38}) for homogeneous elasticity $\beta$ and net cross-section area $S_0$ of vessels, one would find the wave propagation speed $c=\sqrt{\beta S_0}$. Substituting solution~(\ref{eq38}) into equation~(\ref{eq21}), one can obtain the flux
\begin{equation}
q(\xi,t)=q(\xi_0,t)
 +\int_{\xi_0}^\xi\frac{f^{\prime}d\xi_1}{\beta(\xi_1)}\,,
\label{eq310}
\end{equation}
where the prime denotes the derivative of a function with respect to its argument.

As the first step of analysis of solutions (\ref{eq38}) and (\ref{eq310}), we qualitatively consider the issue of the phase shift between the waves of volume and flux.

\subsection{Qualitative analysis of the phase shift: wave extrema}
Let us consider propagation of a solitary wave through the blood-vascular system, choosing the reference point $\xi_0$ behind the wave so, that $q(\xi_0,t)=q_0(\xi_0)$. Physiologically, the location of point $\xi_0$ can be chosen immediately after the heart. At certain point $\xi$, where
\[
\left.\frac{\partial\sigma_1}{\partial t}\right|_\xi=0\,,
\]
differentiating equation~(\ref{eq310}) with respect to time, one finds
\[
 \left.\frac{\partial q}{\partial t}\right|_\xi
 =-\int_{\xi_0}^\xi\frac{f^{\prime\prime}d\xi_1}{\beta(\xi_1)}
 =\left.f^{\prime}(\xi_1)\sqrt{S_0(\xi_1)/\beta(\xi_1)}\right|_\xi^{\xi_0}
 +\int_{\xi_0}^\xi f^\prime
 \frac{d}{d\xi_1}\!\left(\sqrt{\frac{S_0(\xi_1)}{\beta(\xi_1)}}\right)d\xi_1=0\,,
\]
because where $\partial\sigma_1/\partial t=0$, $f^\prime$ also becomes zero (which can be explicitly seen from the time-derivative of expression (\ref{eq310})), and the reflectionlessness condition~(\ref{eq37}). For the case of a train of waves, the situation is the same.

Thus, for inhomogeneous  parameters of the system, the extrema of flux and volume are attained simultaneously (figure~\ref{fig1}). Moreover, one can notice that this property is accurately preserved even for a transient processes, when system parameters are changing with time (in figure~\ref{fig1}b after approximately 90 sec.\ a deep inspiratory gasp manoeuvre begins).

\subsection{The relation between measurable signals}
Let us now derive the relation between measurable signals of the blood volume and flux.
From equation~(\ref{eq38}), one can find
\[
\beta(\xi_1)\,\sigma_1(\xi_1,t)=\beta(\xi)\,\sigma_1(\xi,t+\tau(\xi,\xi_1))
\]
and evaluate the integrand of equation~(\ref{eq310}) from a measurable signal $\sigma_1(\xi_\mathrm{m},t)$ (where $\xi_\mathrm{m}$ is fixed to the observation site, and $\sigma_1(\xi_\mathrm{m},t)$ is measured as a function of time);
\[
\frac{f^\prime\Big(
 \int^{\xi_1}[\beta(\xi_2)\,S_0(\xi_2)]^{-1/2}d\xi_2 -t\Big)}{\beta(\xi_1)}
 =-\dot\sigma_1(\xi_1,t)=-\frac{\beta(\xi_\mathrm{m})}{\beta(\xi_1)}\dot\sigma_1\big(\xi_\mathrm{m},t+\tau(\xi_\mathrm{m},\xi_1)\big)\,,
\]
where the dot stands for the time-derivative. Hence, equation~(\ref{eq310}) takes the form
\begin{equation}
q(\xi_\mathrm{m},t)=q(\xi_0,t)
 -\beta(\xi_\mathrm{m})\int\limits_0^{\tau(\xi_\mathrm{m},\xi_0)}
 \dot\sigma_1(\xi_\mathrm{m},t+\tau)\sqrt{\frac{S_0[\xi_1(\tau)]}{\beta[\xi_1(\tau)]}}\,d\tau\,,
\label{eq311}
\end{equation}
where $\xi_1(\tau)$ is implicitly determined by equation~(\ref{eq39}) or, which is the same but can be more convenient technically in some situations, by the problem
\[
\xi_1(\tau=0)=\xi_\mathrm{m}\,,\qquad
d\xi_1=-\sqrt{\beta(\xi_1)\,S_0(\xi_1)}\,d\tau\,.
\]

For the time instances, {\it when the blood flow and the vessel walls near the heart are at the ground state ({\it i.e.}, between heart pulses)}, one can set $q(\xi_0,t)=q_0(\xi_0)$ in equation~(\ref{eq311}) and find the relation involving only measurable signals $q(\xi_\mathrm{m},t)$ and $\sigma_1(\xi_\mathrm{m},t)$;
\begin{equation}
q(\xi_\mathrm{m},t)=q_0(\xi_0)
 -\beta(\xi_\mathrm{m})\int\limits_0^{\tau(\xi_\mathrm{m},\xi_0)}
 \dot\sigma_1(\xi_\mathrm{m},t+\tau)\sqrt{\frac{S_0[\xi_1(\tau)]}{\beta[\xi_1(\tau)]}}\,d\tau\,.
\label{eq312}
\end{equation}

\section{Conclusion}
We have analytically demonstrated, that elastic waves of volume and flux in the vessels of the blood-vascular system attain extrema simultaneously even in the case where the inhomogeneity of parameters is significant. The experimental observations also suggests nearly no shift of principal extrema of PPG- and LDF-signals for unsteady conditions (see figure~\ref{fig1}b). The observed phase shift between the extracted $1\,\mathrm{Hz}$-components of PPG- and LDF-signals~\cite{Podtaev-Mizeva-Alan-2012}, which is not detected for steady conditions and appears for unsteady ones, is presumably associated with an integral change in the pulse forms.

The relation between waveforms of volume and flux has been derived (equations~(\ref{eq38}) and (\ref{eq310}) {\it or} equations~(\ref{eq311}) and (\ref{eq312}), where the latter equation is for specific time instances). These relations provides opportunity for extraction of diverse information on the variation of the system parameters (elasticity and inner diameter of vessels along the network) in response to different external conditions and functional tests from the readings of PPG and LDF.

For the system under consideration the spacial inhomogeneity of parameters is essential as they change by two orders of magnitude along the system and the spatial extent of the wave pulses is commensurable to the characteristic scale of the parameter inhomogeneity. A waveguide with such a strong inhomogeneity of parameters would be generally nearly impenetrable for waves, reflecting the major part of the energy of incident waves. In practice, the system seems to be optimised from the view point of reflectionlessness; the reflection in a healthy state is possibly minimal. The approximation of the reflectionlessness of the system has allowed us to derive the reported relations between waveforms of blood volume and flux.

As an important implication of the reflectionlessness conjecture, one can expect that an increasing pathological variation of the vessel inner diameters and elasticity properties should result in the growing violation of the reflectionlessness conditions and thus also in an increasing stress impact on both the heart and large vessels by reflected waves.

\ack{The work has been financially supported by RFBR (grant no.\ 17-41-590560 r\_a).}

\end{document}